\documentclass[pre,twocolumn]{revtex4}
\usepackage{graphicx}
\usepackage{amsmath}

\begin{document}

\title{Evolution of Canalizing Boolean Networks}
\author{A. Szejka, B. Drossel}
\affiliation{Institut f\"ur Festk\"orperphysik, TU Darmstadt, Hochschulstrasse 6, 64289 Darmstadt, Germany}
\date{\today}

\begin{abstract}
Boolean networks with canalizing functions are used to model gene regulatory networks. In order to learn how such networks may behave under evolutionary forces, we simulate the evolution of a single Boolean network by means of an adaptive walk, which allows us to explore the fitness landscape. Mutations change the connections and the functions of the nodes. Our fitness criterion is the robustness of the dynamical attractors against small perturbations. We find that with this fitness criterion the global maximum is always reached and that there is a huge neutral space of 100\% fitness. Furthermore, in spite of having such a high degree of robustness, the evolved networks still share many features with ``chaotic'' networks.
\end{abstract}
\pacs{89.23.Kg, 89.75.Hc, 87.14.Aa}
\maketitle

\section{Introduction} \label{int}
In a gene regulatory network a protein, which is encoded by a gene, can regulate the expression of one or several other genes, usually in cooperation with other proteins. This results in a complex network of interacting units. In 1969, Stuart Kauffman was the first one to model such a gene regulatory network by means of a random Boolean network \cite{kau1969a,kau1969b}. In this simple model each gene $i$ can be in two different states, $\sigma_i=1$ or 0. This means that the gene is either expressed or not expressed. Each gene is represented by a node and each interaction by a directed connection between two nodes. 
Each node $i$ receives input from $K_i$ randomly chosen other nodes, and its state at time step $t$ is a function of the states of its input nodes at time step $t-1$,
\begin{equation}
\sigma_i(t) = f_i[\sigma_{i_1}(t-1), \sigma_{i_2}(t-1), ..., \sigma_{i_{K_i}}(t-1)]
\end{equation}
Starting from any of the $2^N$ possible network states, where $N$ is the number of nodes, the network eventually settles on a periodic attractor. Usually, there are different attractors with different basins of attraction (i.e., the set of states leading to and lying on the attractor) and attractor lengths (i.e., the number of states on the attractor).

Random Boolean networks can be in two phases, the frozen and the chaotic phase, that differ greatly in the dynamical behavior of the networks. A network is said to be in the frozen phase if a perturbation at one node propagates during one time step on an average to less than one other node. In the chaotic phase the difference between two almost identical states increases exponentially fast, because a perturbation propagates on an average to more than one node during one time step. Kauffman argued that a real genetic regulatory system should be \textit{critical}, that is on or near the boundary between the two phases, thus being both stable and evolvable \cite{kau1993}.

As pointed out in \cite{bor2005}, the Boolean idealization seems to be suitable to model the overall dynamical properties of gene regulatory networks. The segment polarity network of Drosophila melanogaster \cite{alb2003} and the yeast cell-cycle network \cite{li2004}, for example, were modeled using Boolean dynamics of the genes, and both models show attractors that agree with the biological sequence of events. Additionally, the analysis of the data for over 150 gene regulatory systems of eukaryotes revealed a strong bias towards canalizing functions (defined below) \cite{har2002}. However, while gene regulatory networks may show the same type of dynamics as random Boolean networks, the structure of gene regulatory networks is very different and far from random. They have attractors that are far more robust and have much larger basins of attraction than random Boolean networks. Biological networks are shaped by their evolutionary history, and they are designed to execute certain tasks. Neither of these features is included in the original models. 

In this article, we want to address the first of these two issues and investigate how initially random Boolean networks change under the evolutionary forces of mutation and selection. The simulation of evolution in Boolean networks has a long history, and several approaches have been taken. In \cite{Kauffman86}, networks with connectivities of $K = 2$ and $K = 10$ are evolved by creating mutants through the rewiring of connections or the changing of bits of Boolean functions. For every network, attractors are found by letting the network run from a predetermined number of random initial states, and these attractors are compared to a ``target'' state. The fitness of each network is defined as the mean Hamming distance of the target state to the attractor closest to it. The fittest network is used to generate the mutants of the next generation. During evolution, fitness approaches an asymptotic value that is less than the global maximum.

In \cite{Bornholdt98}, the evolution of the connectivity of a single \linebreak Boolean network is studied. A random Boolean network with $K = 1$ is started from a random initial state. When it reaches an attractor, a daughter network is created by adding and/or removing a connection at random. If the daughter network reaches the same attractor after starting from the same initial state the simulation is continued with the daughter network, otherwise with the mother network. The evolution of the network connectivity shows ``punctuated equilibrium'', as observed in the fossil record, with the network ``species'' with high $K$ values having extremely long life times. The continuation of this work with threshold networks \cite{Bornholdt00} reveals that the evolved networks have shorter attractors and a larger frozen component than random networks.

In \cite{Stern99}, the influence of noise on the evolution of Boolean networks is studied. A population of Boolean networks is evolved under the selection criterion of matching a given target pattern while it is subjected to different kinds of noise. Mutations consist in redirecting a randomly chosen link or in changing a randomly chosen update function. Noise exclusion as well as ``noise imprinting'' is observed. Noise in the fitness measurement makes adaptive evolution impossible if the noise level exceeds a critical value.

In \cite{Paczuski00}, an economic network is modeled where the nodes represent competing agents, and a game determines the process of evolution. In every time step, the winning nodes are those that have the same state as the minority of the nodes. The node that has lost the game most often during a certain time is assigned a new random Boolean function. The network evolves to a stationary state ``at the edge of chaos''. Further studies of the model \cite{Bassler04} showed that the evolved networks are highly canalized. Other models that evolve to a critical state are studied in \cite{Luque01,BornholdtRohlf00,Liu06}.

In our work reported in this paper, we simulated the evolution of a single canalizing Boolean network (as opposed to the simulation of a whole population of networks as for example in \cite{Stern99}). The mutations affect the connections and the update functions of the nodes. Compared to the simulations in \cite{Kauffman86,Bornholdt98,Bornholdt00,Stern99,Paczuski00,Bassler04,Luque01,BornholdtRohlf00,Liu06}, fitness can be changed by more types of mutations in our model. We let the networks evolve freely under a biologically motivated fitness criterion without imposing any ``target properties'' to find network topologies that are evolutionary robust. The fitness criterion is robustness against small perturbations. We interpret robustness as the ability of a system to maintain function in the face of perturbations/noise (in contrast to mutational robustness considered for example in \cite{Bornholdt98,Bornholdt00}). Robustness is of great importance in biology as a cell must continue to function and to pass on its genetic material in the face of fluctuations e.g. of the protein concentrations or of the nutrient levels. We simulated an evolutionary process that is a so-called adaptive walk. This is a hill climbing process that leads to a local fitness maximum and thus can yield insight into the fitness landscape of a system. Our main finding is that the maximum possible fitness value is always reached during this process, and that there is a huge plateau with this fitness value that spans the network configuration space.

In the next section, we define the set of canalizing functions we used in our simulations. Then, the algorithm for the adaptive walk is described in section \ref{compsym}. In section \ref{res}, the results of the simulations are given. The last two sections include a discussion of these results and a final summary.

\section{Canalizing Functions} \label{canfunc}
A function is called canalizing if at least one value of one input can determine the output of the function, independently of the other inputs. This means that the output is fixed when the canalizing variable takes the canalizing value. Here we choose the set of canalizing functions used by Moreira and Amaral \cite{mor2005}, where frozen functions are a special case of canalizing functions. Four classes of functions are distinguished: 
\begin{align}
	F(\sigma_1, \sigma_2, ...)& = \sigma_1 \hspace{0.2 cm} OR \hspace{0.2 cm} G(\sigma_2, ...) \tag{2.1} \\ 
	F(\sigma_1, \sigma_2, ...)& = (NOT \hspace{0.1 cm}\sigma_1) \hspace{0.2 cm} AND \hspace{0.2 cm} G(\sigma_2, ...)\tag{2.2} \\ 
	F(\sigma_1, \sigma_2, ...)& = (NOT \hspace{0.1 cm}\sigma_1) \hspace{0.2 cm} OR \hspace{0.2 cm} G(\sigma_2, ...)\tag{2.3} \\
	F(\sigma_1, \sigma_2, ...)& = \sigma_1 \hspace{0.2 cm} AND \hspace{0.2 cm} G(\sigma_2, ...)\tag{2.4}
\end{align}
where $\sigma_1$ is the canalizing variable and $G$ is a random Boolean function that depends on the remaining variables.

For classes (2.1) and (2.2), the canalizing value is 1, classes (2.3) and (2.4) have the canalizing value 0. The canalized value (the value the function yields when the canalizing variable takes the canalizing value) is 1 for classes (2.1) and (2.3) and 0 for classes (2.2) and (2.4).

In their paper Moreira and Amaral determine the conditions under which networks with canalizing functions are critical. If all functions mentioned above are chosen with equal probability, the critical number of inputs per node is $K_c = 3$.

Kauffman et al. found that networks with canalizing rules are remarkably stable compared to those with random Boolean functions \cite{kau2003}.

\section{Computer Simulations}\label{compsym}
The adaptive walk starts with a randomly created network. The fitness of this network is determined, and then a mutation is performed in the network. The mutation is accepted if it does not lower the fitness. \textit{Neutral} mutations are those that do not change the fitness value. Then the next mutation is attempted. This procedure is continued until a certain stopping condition is satisfied. In detail the algorithm is defined as follows:
\begin{enumerate}
\item A network with $N$ nodes is generated. Each node is assigned a random initial value (0 or 1) and $K$ randomly chosen input nodes. (The number of inputs is the same for all nodes in the beginning, but later on different nodes can have different $K$ values.) The first input of each node is the canalizing one. Then a random Boolean function $G$ with $K-1$ inputs is generated for each node by choosing $0$ or $1$ with the same probability as output for every input combination. Last, one of the four classes of canalizing functions (2.1) to (2.4) is assigned to each node with equal probability. We thus have obtained a network \textit {realization}.
\item To determine the fitness (that is the robustness) of the network, three steps are necessary. First the network is updated according to equation (1) until it reaches an attractor. Then the value of each node is flipped one after the other, and it is counted for how many of these flips the network returns to the same attractor. This can happen at most $N$ times. The fitness value is the percentage of times the dynamics return to the given attractor after flipping a node.
\item Then a mutation is performed at a randomly chosen node. Each of the following four mutations occurs with probability 1/4:
\begin{enumerate}
\item a connection is added; (The maximum number of inputs to a node is limited to $K_{max} = 10$ due to computational restrictions.)
\item a connection is deleted;
\item a connection is redirected;
\item the canalizing function is changed.
\end{enumerate}
Mutations (b), (c) and (d) fail if $K = 0$ for the chosen node. Mutation (a) fails if $K = 10$. In these situations, it is attempted up to $N^2$ times to find a node for which the mutation does not fail. If no such node is found, a new mutation is chosen with equal probability from the four possible ones. Note that whenever a connection is added or deleted, the random Boolean part of the update function has to be newly generated. If a node looses the connection to its canalizing input the next input is chosen to be canalizing.\\
During the first simulation runs, we worked with a value $K_{max} = 7$, but as all the networks reached this number of inputs during evolution, we then chose $K_{max} = 10$. Because the simulations become slower with increasing $K$, we did not go beyond this value.
\item The adaptive walk is stopped when one of the following three events occurs:
\begin{enumerate}
\item a certain number of attempted mutations is exceeded and no mutation was accepted; (This would happen for example if the network was on a local maximum in the fitness landscape and all mutations decreased the fitness. Since the simulations show that the networks never get stuck at local maxima, the threshold can be chosen such that this stopping criterion is never fulfilled.)
\item trying to add a link failed $N$ times because $K = K_{max}$;
\item a certain number of accepted mutations is reached. (This number was 100,000 or 200,000 in the simulations evaluated below.)
\end{enumerate}
\end{enumerate}
This is the basic algorithm. Changes to it are mentioned where the according variations are discussed in the results section.

\section{Results} \label{res}
We have simulated adaptive walks for networks with $N =$ 20, 30, 50, 70, 80 and 100 nodes and for initial connectivities of $K_{ini} =$ 1, 2, 3, 4, 5 and 6.

In the next two subsections we study the path taken by the networks until they reach maximum fitness. From section \ref{neutev} on we examine also the evolution of the networks after reaching maximum fitness.
In the figures that deal with the evolution of the networks to the fitness maximum, every data point is averaged over at least 5000 adaptive walks. Exceptions are the simulations for $N = 100$, $K = 5$ with only 3600 runs, the networks larger than 100 nodes in Fig. 1, and the simulations without neutral mutations that needed less runs (Fig. 3). In the figures that deal with the evolution of the networks after they have reached maximum fitness every curve corresponds to one adaptive walk.

Therewith the first remarkable result is already stated: all networks always reach maximum fitness, independently of the initial conditions.

\subsection{Initial Fitness}
\begin{figure}[t]
\includegraphics[angle = -90, width = \columnwidth]{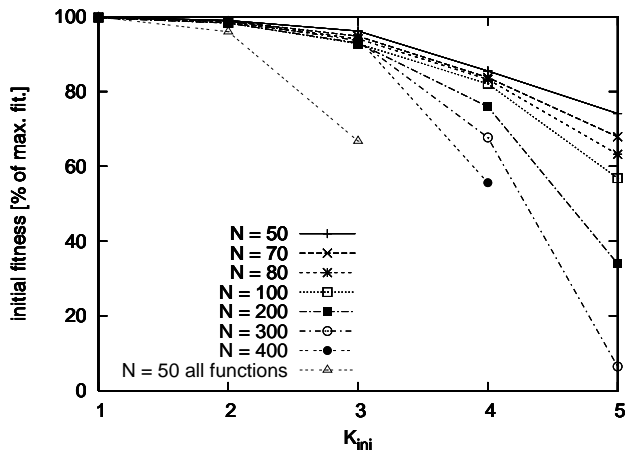}
\caption{The initial fitness of networks with different $N$ and $K_{ini}$.
For comparison the initial fitness of a random network with all Boolean functions is plotted for $N =$ 50 and $K_{ini} =$ 1 to 3.}
\label{inifit}
\end{figure}
Fig. \ref{inifit} shows the fitness of the networks generated at the beginning of the simulations. One can see that the initial fitness decreases with increasing $K_{ini}$ and $N$. For comparison the fitness of a random Boolean network with all possible Boolean functions (and not just the canalizing ones), $N~=$~50 and $K_{ini} =$ 1, 2 and 3 is plotted. (Simulations with higher initial $K$ are too time-consuming because of the extremely long attractors.) One can see that networks with all possible Boolean functions are less stable than those with only canalizing functions. Furthermore, there is a clear decline of the fitness at $K_{ini} =$ 2, beyond which this kind of network becomes chaotic. For canalizing networks, the critical number of inputs per node is $K_c =$ 3, and the curves for the largest three $N$ values intersect at this point. However the network sizes considered in the rest of this paper (up to $N =$ 100) are too small to see a clear phase transition.

\begin{figure}[t]
\hspace{-0.25cm}\includegraphics[angle = -90, width = \columnwidth]{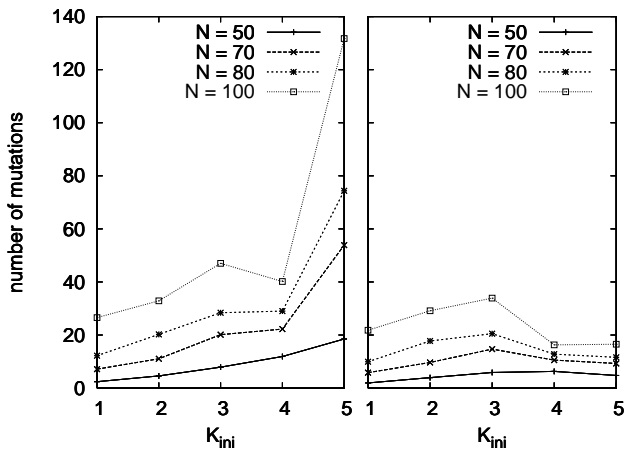}
\caption{The number of attempted mutations (left) and accepted mutations (right) needed until the network reaches 100\% fitness, as function of the initial connectivity.}
\label{mutnew}
\end{figure}

\subsection{Length of Adaptive Walk}
In this subsection we study the path taken by the networks until they reach the global fitness maximum. As stated before, all networks reach maximum fitness, local maxima are not found. In the figure caption, the \textit{attempted} mutations comprise all mutations that were performed, whether they were accepted or not. The \textit{accepted} mutations are those that increase or do not change the fitness, that is, the ones that determine the path of the networks through the fitness landscape. Maximum fitness is reached after only a few evolutionary steps for all network realizations. Fig. \ref{mutnew} (right side) shows that the number of 40 accepted mutations is never exceeded for the network sizes considered. This number is so small that the $K$ values of the networks do not change much until the arrival at the fitness maximum. In general, for larger $N$ more mutations are necessary to reach maximum fitness. With increasing $N$, both sets of curves appear to develop a local maximum at the critical value $K_{ini}=3$. It seems that networks that start at the critical point have a longer path to the global peak of the fitness landscape than mostly frozen or slightly chaotic networks. While the number of attempted mutations (left curve) increases again for $K_{ini}$ values larger than 4, this is not so evident for the number of accepted mutations. The last two points of the curve for $N=100$ suggest that for larger $N$ there might be also an increase in the number of accepted mutations with increasing $K_{ini}$. 

\begin{figure}[t]
\includegraphics[angle = -90, width = \columnwidth]{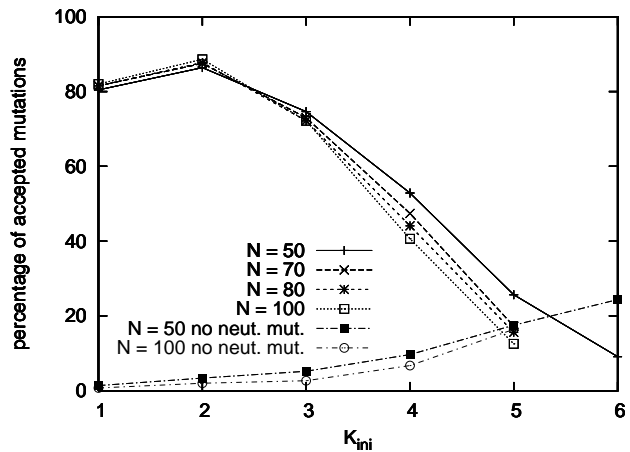} 
\caption{The percentage of accepted mutations until the arrival at the global maximum. The upper four curves were obtained with our usual simulations. The lower two curves were obtained with modified simulations where neutral mutations were not accepted but only mutations that increased the fitness.}
\label{acc_mut_percent}
\end{figure}

In Fig. \ref{acc_mut_percent}, the ratio of accepted to attempted mutations is shown for different network sizes and connectivities. For $K_{ini} >$ 2 finding mutations that are accepted is more difficult when the networks are larger, and the percentage of accepted mutations shrinks with growing initial connectivity. The comparatively low percentage of accepted mutations at $K_{ini} =$ 1 can be easily understood since the initial fitness is already close to 100\%. Fig. \ref{acc_mut_percent} shows also two curves for $N =$ 50 and 100 from simulations where neutral mutations (mutations that do not change the fitness) were not accepted. Thus the total number of accepted mutations is smaller in this case. In this respect the path to the global maximum is shorter. But since a higher number of attempted mutations is needed, the adaptive walk is less efficient for $K_{ini}<5$, as one can see in Fig. \ref{acc_mut_percent}. The percentage of accepted mutations is much smaller for adaptive walks without neutral mutations.
More evolutionary time is needed to take the networks to the global maximum. The situation changes for larger $K_{ini}$. For networks with $N =$ 50 the evolutionary process without neutral mutations becomes more efficient at $K_{ini} =$ 6, for networks with $N =$ 100 this happens already at $K_{ini} =$ 5. This means that at these points the number of attempted mutations increases drastically for simulations with neutral mutations while it decreases for simulations without neutral mutations. Apparently, allowing the network to undergo neutral mutations drives it into regions in network space where there are more neutral directions and less uphill neighbors.

\begin{figure}[t]
\includegraphics[angle = -90, width = \columnwidth]{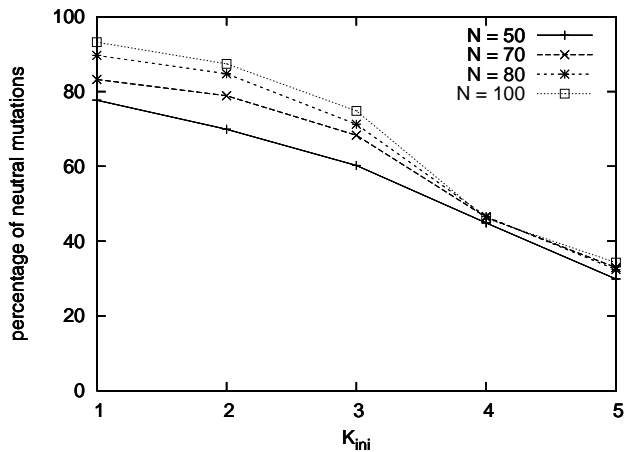}
\caption{The percentage of neutral mutations among the accepted mutations until the networks reach 100\% fitness.}
\label{neut_mut}
\end{figure}
\begin{figure}[t]
\includegraphics[angle = -90, width = \columnwidth]{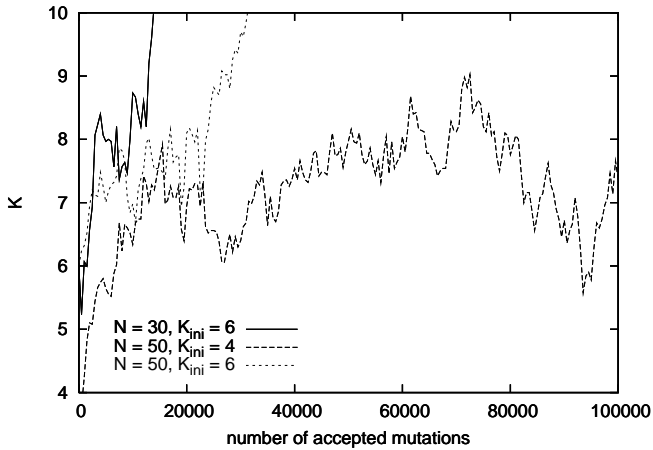}
\caption{The average connectivity $K$ as function of time (measured in accepted mutations) for different initial connectivities and different network sizes.}
\label{a}
\end{figure}

\subsection{Neutral Evolution} \label{neutev}
When looking at the number of neutral mutations that occur until the networks reach maximum fitness, one can see that their percentage decreases with increasing initial $K$ (Fig. \ref{neut_mut}). Furthermore, the percentage of neutral mutations is larger for larger networks.
After reaching maximum fitness, the adaptive walk continues via neutral mutations. Thus the global maximum of the fitness landscape is not an isolated point from which every path leads downhill, but a far-ranging plateau, on which the networks can move in many directions. Fig. \ref{a} shows the evolution of the average connectivity over 100,000 evolutionary steps for different network sizes and initial connectivities. (The initial connectivity is actually of little importance as the networks ``forget'' about it after a few thousand mutations.) It appears that the networks perform a random walk through $K$-space and can reach in principle every possible $K$ value. The adaptive walk could go on forever if there were no stopping conditions.

At the end of the adaptive walk the in-degree and out-degree distributions are either both Poissonian, or the in-degree distribution is a delta function because all nodes have reached the maximum possible $K$.

\subsection{Variations}
This section deals with variations of the basic algorithm. The variations concern the probabilities for the occurrence of the different mutations, the fitness criterion and the initial distribution of $K$ values. At the end of this section, the evolution of random Boolean networks with all possible Boolean functions is considered for comparison.

\begin{figure}[t]
\includegraphics[angle = -90, width = \columnwidth]{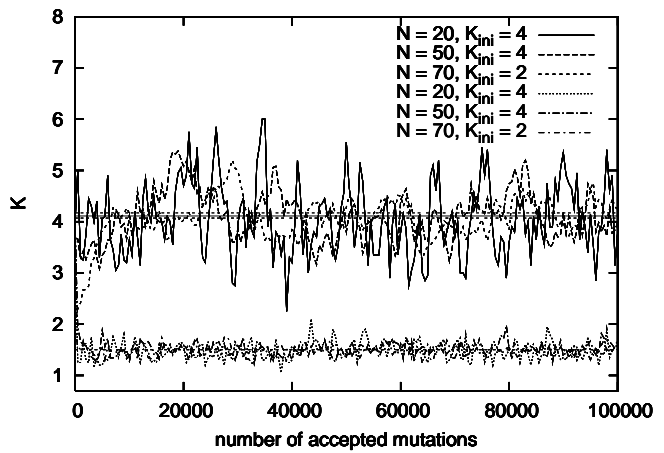}
\caption{The $K$ values of six different network realizations with two different probability distributions for the mutations as function of time. The horizontal lines indicated the time average of $K$ for each curve.}
\label{ame}
\end{figure}

\subsubsection{Bias to Deletion}
We modified the probability distributions for mutations such that the probability to add a connection was smaller than the probability to delete a connection. This leads to the stabilization of the networks at a certain $K$ value. In Fig. \ref{ame} we show the $K$ value over a time period of 100,000 mutations. For the curves with the higher average $K$ the probability to delete a link was 0.275 and the probability to add a link was 0.225. For the curves with the lower average $K$ these values were 0.375 and 0.125 respectively. The probabilities for the two other mutations remained unchanged. 
During the adaptive walk the connectivity fluctuates around a mean value that appears to be independent of network size. Assuming that the networks have reached the stationary state after about 10,000 accepted mutations, the mean $K$ values of the upper curves (smaller bias) are found to be 4.07, 4.17 and 4.11 for network sizes of $N~=$~20, 50 and 70 respectively, with $\Delta K / \langle K \rangle$ being 0.27, 0.17 and 0.13. The lower curves (larger bias) have mean values of $\langle K \rangle$~=~1.49, 1.50 and 1.51 respectively with $\Delta K / \langle K \rangle =$ 0.18, 0.12 and 0.11, which is of the same order as for smaller bias.

\begin{figure}[t]
\includegraphics[angle = -90, width = \columnwidth]{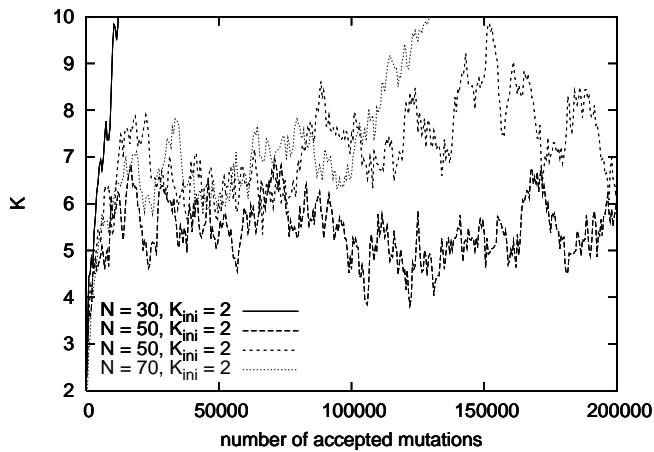}
\caption{Time evolution of $K$ over 200,000 mutations for four different realizations with the restriction that attractor length must not become longer.}
\label{aFP}
\end{figure}
\begin{figure}
\includegraphics[angle = -90, width = \columnwidth]{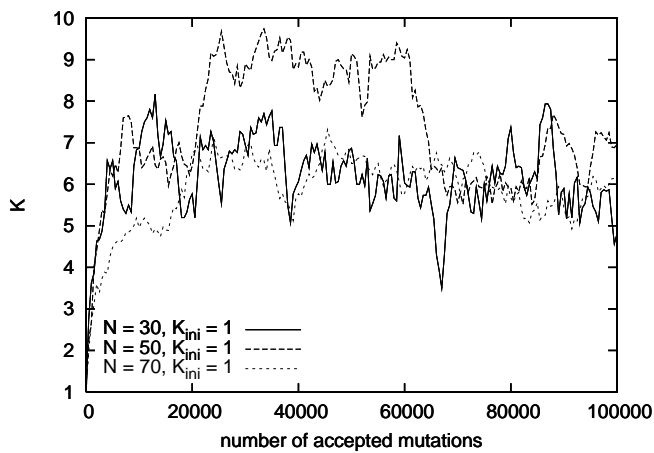}
\caption{Time evolution of $K$ over 100,000 mutations for three different realizations with the restriction that the way back to the attractor after a perturbation should never be longer than two transient states.}
\label{ats3}
\end{figure}
\subsubsection{Stricter Robustness Criteria}
For higher $K$ values, the lengths of the attractors and of the paths back to the attractor after a perturbation are longer. We therefore explored two stricter robustness requirements. First, we accepted only mutations for which the attractor lengths did not grow. Second, we included in the fitness criterion the condition that there should be at most two transient states on the way back to the attractor. The first modification leads to an attractor of length 1 within a few evolutionary steps, which corresponds to a fixed point. Even with this restriction the networks manage to reach every possible $K$ value during evolution (Fig. \ref{aFP}). The evolution of the $K$ value for the second modification is shown in Fig. \ref{ats3}. The $K$ value shows again large fluctuations in time. The effect of both constraints is similar: The fluctuations of $K$ appear smaller than before (cf{.}~Fig. \ref{a}), and the attractors become more stable in the sense that more initial states lead to the same attractor (see Sec. \ref{rob}). The two constraints appear to cause each other: When shorter attractors are selected for, the number of time steps needed to return to the attractor also become fewer and vice versa.

From a further comparison of Figs. \ref{a}, \ref{aFP} and \ref{ats3} one can draw the conclusion that if the networks exceed a certain $K$ value, mutations that increase $K$ further are more efficient.

\begin{figure}[t]
\includegraphics[angle = -90, width = \columnwidth]{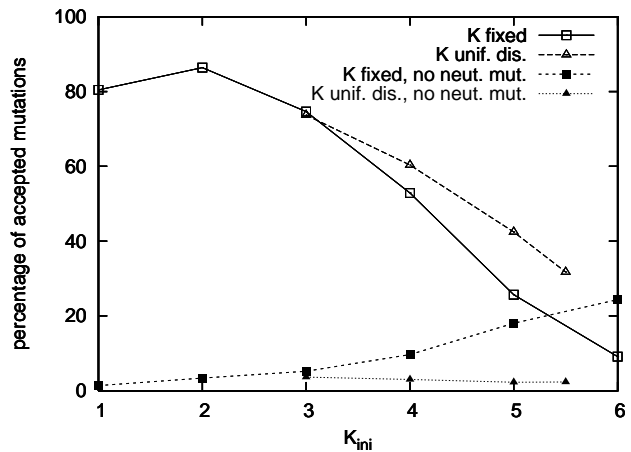}
\caption{Percentage of accepted mutations for networks with $N = 50$ and a uniform distribution of initial $K$ values for simulations with (open symbols) and without (filled symbols) neutral mutations. For comparison the corresponding curves for fixed initial $K$ values from Fig. \ref{acc_mut_percent} are shown again.}
\label{Kdis}
\end{figure}
\subsubsection{Uniform Distribution of Initial $K$ Values}
The simulations considered so far started with networks where all nodes had the same number of inputs. As the path of the networks to the global fitness maximum depends crucially on the initial connectivity, we also studied adaptive walks where the $K$ values were uniformly distributed between 1 and $K_{max}$ in the beginning. We did simulations with $N = 50$ and $K_{max} =$ 5, 7, 9 and 10, so that the mean initial $K$ was 3, 4, 5 and 5.5 respectively. Compared to the earlier simulations, the networks need more mutations to reach the global maximum. These are still relatively few mutations, so that the mean connectivity does not change much. In Fig.~\ref{Kdis} one can see that for simulations where $K$ values were uniformly distributed in the beginning, more mutations are accepted if neutral mutations are allowed. Also the percentage of neutral mutations among the accepted mutations is higher (not shown). In contrast, during simulations without neutral mutations it was much more difficult to find mutations that are accepted. We conclude that networks with a broader initial distribution of $K$ values find more neutral directions in the fitness landscape at the cost of having a longer path to 100\% fitness. As expected, the initial $K$ distribution does not affect the long term evolution after networks have reached 100\% fitness.

\subsubsection{All Boolean Functions}
\begin{figure}[t]
\includegraphics[angle = -90, width = \columnwidth]{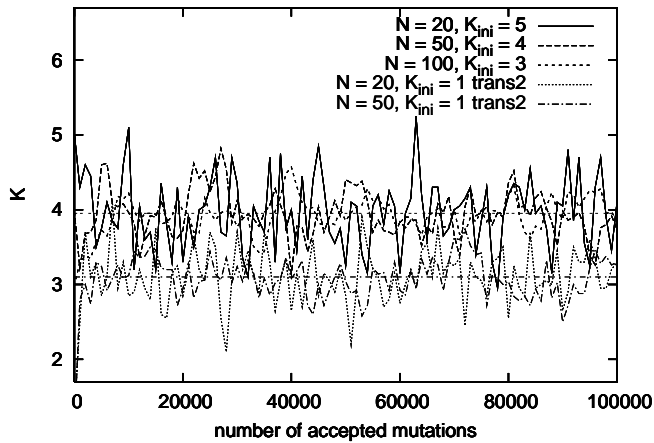}
\caption{Time evolution of $K$ over 100,000 mutational steps for networks with all Boolean functions. The $K$ value fluctuates around a mean of $K = 3.95$ for all network sizes. With the restriction of at most two transient states after perturbation, this value decreases to $K = 3.10$. The horizontal lines show these average $K$ values.}
\label{aBtrans2}
\end{figure}

If not only canalizing functions but all Boolean functions are allowed, the $K$ value fluctuates after about 10,000 mutations around a mean value that is independent of the network size (Fig. \ref{aBtrans2}). This value is $3.95 \pm 0.05$ for $N = 20$, $3.94 \pm 0.03$ for $N = 50$ and $3.95 \pm 0.02$ for $N = 100$. When additionally the above-mentioned constraint of at most two transient states is introduced, the average $K$ value shrinks to $3.11 \pm 0.04$ for $N = 20$ and $3.10 \pm 0.03$ for $N = 50$. We also found that the proportion of the different Boolean functions does not change during evolution, and therefore the networks still have $K$ values characteristic of the chaotic regime.

\section{Discussion} \label{dis}
\subsection{Robustness and Chaos} \label{rob}
As stated in the previous section, the networks that were evolved using all Boolean functions have $K$ values characteristic of the chaotic regime. The networks with only canalizing functions have $K$ values higher than $K_c$ most of the time. In both cases, the networks show features of real genetic networks: they are stable and evolvable.
In order to understand how the networks can be so robust in spite of high $K$ values, we carried out some investigations on the evolved networks.
First, we tested the robustness $r$ of the evolved networks; $r$ is the probability that the state of a node changes when one of its inputs changes \cite{Kauffman04}. The evolved networks are not more robust than the initial networks according to this criterion.
Second, we examined the degree of canalization of the evolved networks, but neither the number of canalizing inputs nor the degree of nesting of the canalizing functions differ from a random distribution.
Finally, we examined the state space of the evolved networks to study the stability of the attractors. For networks with $N=$ 20, we explored the entire state space, for networks with higher $N$ we considered 10,000 or 100,000 initial states. We found that the evolved networks have an attractor that attracts at least 90\% of all initial states. (The same holds for networks with all Boolean functions.) This means that most perturbations, no matter how far away from the attractor they carry the network, have to lead back to the attractor eventually. The networks evolved with the stricter robustness criteria are even more stable in this sense. 

The selection criterion used in this paper leads to networks that show stable behavior under perturbations, and at the same time they display properties of chaotic networks. The relatively long attractors found for networks evolved with the original selection criterion are typical for the chaotic or the critical regime. We evaluated the activity of the nodes on the attractors and found that approximately half of the nodes are blinking, which means that only half of the nodes are frozen. That is also a characteristic of chaotic networks.

\subsection{Earlier simulations of network evolution}
Let us compare some of our results with results from earlier simulations of the evolution of Boolean networks (see Sec. \ref{int}).
In \cite{Kauffman86}, networks never reach maximum fitness. In contrast, the networks in our simulations always reach maximum fitness even under various constraints. This difference may be due to the fundamentally different fitness criteria. In \cite{Kauffman86}, the fitness criterion imposes a ``goal'' on evolution. Another possible explanation might be the larger variety of mutations in our simulations. In \cite{Kauffman86}, $K$ did not change.
The work with which our study overlaps most is \cite{Bornholdt98,Bornholdt00}, where the evolution of Boolean networks was performed by selecting for robustness, although in those papers the robustness is phenotypic robustness against mutations. Similarly to what we found, the evolutionary processes in \cite{Bornholdt98,Bornholdt00} lead to networks that do not show more chaos when the connectivity grows, but the evolved networks have shorter attractors than the initial networks, while approximately half of the nodes are frozen.

\section{Summary} \label{sum}
In this paper we analyzed the fitness landscape of a genetic regulatory network model. We showed that the networks can reach the maximum possible fitness value when the selection criterion is robustness of attractors against perturbations. Moreover the networks need only a few evolutionary steps to arrive at this maximum. The number of steps seems to be less for networks starting in the frozen or in the chaotic phase than for initially critical networks.
Everywhere in the fitness landscape there exist paths going uphill towards the plateau of maximum fitness. When neutral mutations are allowed, the adaptive walk becomes more efficient when the initial connectivity is small. The plateau of maximum fitness spans the network space, and networks that have reached the fitness maximum can move through this plateau by neutral evolution.
Over long evolutionary time scales, none of the mutations is favored. This means that on average the proportion of a certain type of mutation among the accepted mutations is the same as its proportion among the attempted mutations. 
The stability of the evolved networks is due to a state space containing an attractor with a huge basin that comprises at least 90\% of the possible initial states.

Applied to real genetic networks, our results would mean that their attractors are already relatively stable because of the canalizing functions and that they can become even more stable within short evolutionary time periods if selection is favoring robustness against perturbations. After the networks have reached high fitness values, they easily continue to evolve in different directions without having to reduce their fitness. That is, these networks are very evolvable and nevertheless stable without being restricted neither to small connectivities nor to special mutations. The notion introduced by S. Kauffman that such networks should be near a critical point, that is, at the edge of chaos, appears to be not fully appropriate. Our evolved networks, in addition to being robust and evolvable, share otherwise many features with chaotic networks. The simple classification of network dynamics into frozen, critical, and chaotic, might need to be reconsidered in the light of the different types of dynamical behaviors that can be seen in evolved networks. 

We thank Tamara Mihaljev for useful discussions.

\end{document}